\pdfoutput=1

\documentclass[twocolumn,prl,showpacs,amsmath,amssymb,superscriptaddress]{revtex4}

\usepackage{bm}
\usepackage{graphicx}
\usepackage{amssymb}
\usepackage{xcolor}

\begin{document}

\title{Short-range antiferromagnetic correlations in the superconducting state of filled skutterudite alloys Pr$_{1-x}$Eu$_x$Pt$_4$Ge$_{12}$  }

\author{R. B. Adhikari}  
\affiliation{Department of Physics, Kent State University, Kent, Ohio, 44242, USA}

\author{D. L. Kunwar}  
\affiliation{Department of Physics, Kent State University, Kent, Ohio, 44242, USA}

\author{I. Jeon}
\affiliation{Center for Advanced Nanoscience, University of California, San Diego, La Jolla, California 92093, USA}
\affiliation{Materials Science and Engineering Program, University of California, San Diego, La Jolla, California 92093, USA}

\author{M. B. Maple}
\affiliation{Center for Advanced Nanoscience, University of California, San Diego, La Jolla, California 92093, USA}
\affiliation{Materials Science and Engineering Program, University of California, San Diego, La Jolla, California 92093, USA}
\affiliation{Department of Physics, University of California at San Diego, La Jolla, CA 92903, USA}

\author{M. Dzero}
\affiliation{Department of Physics, Kent State University, Kent, Ohio, 44242, USA}

\author{C. C. Almasan}
\affiliation{Department of Physics, Kent State University, Kent, Ohio, 44242, USA}

\date{\today}

\begin{abstract}
Motivated by  current research efforts towards exploring the interplay between magnetism and superconductivity in multiband electronic systems, we have investigated the effects of Eu substitution through thermodynamic measurements on the superconducting filled skutterudite alloys Pr$_{1-x}$Eu$_x$Pt$_4$Ge$_{12}$. An increase in Eu concentration leads to a suppression of the superconducting transition temperature consistent with an increase of magnetic entropy due to Eu local moments. While the low-temperature heat capacity anomaly is present over the whole doping range, we find that in alloys with $x\leq0.5$ the Schottky peaks in the heat capacity in the superconducting state appear to be due to Zeeman splitting by an internal magnetic field. Our theoretical modeling suggests that this field is a result of the short-range antiferromagnetic correlations between the europium ions. For the samples with $x > 0.5$, the peaks in the heat capacity signal the onset of  antiferromagnetic (AFM) ordering of the Eu moments. 
\end{abstract}


\maketitle
\section{Introduction}
Since the pioneering work of 1958 by Matthias and coworkers which unveiled the antagonistic nature of magnetism and conventional superconductivity (SC) \cite{Matthias1958}, research efforts focused on a deeper understanding of  unconventional SC have revealed the coexistence of these competing types of order in a certain region of the phase diagram of unconventional superconductors. While in single-band superconductors, magnetism and conventional superconductivity may co-exist in a fairly narrow region of the material's phase diagram provided the corresponding Curie temperature is lower than the superconducting critical temperature \cite{TheoryOfMetals}, in multiband superconductors the region of co-existence is usually much broader (for the most recent examples of such a situation see, e.g., Ref. \cite{VC_Disorder2011} and references therein).

Filled skutterudite compounds with the chemical formula $M$Pt$_4$Ge$_{12}$ ($M =$ alkaline earth, lanthanide, or actinide) represent an example of an electronic system in which localized $4f$ moments order antiferromagnetically  in the superconducting state \cite{Maple2000}.  The first Pr-based heavy-fermion superconductor PrOs$_4$Sb$_{12}$ has a superconducting critical temperature \linebreak $T_c\simeq 1.85$ K and a normal-state Sommerfeld coefficient \linebreak $\gamma_n\sim 500$ mJ/(mol$\cdot$K$^2$), revealing a rather significant enhancement of the effective mass of the conduction electrons~\cite{Bauer2002,Maple2006}. The related compound PrPt$_4$Ge$_{12}$ has a much higher $T_c\simeq{7.9}$ K and smaller \linebreak $\gamma_n\sim 60$ mJ/(mol$\cdot$K$^2$), corresponding to a moderate enhancement of the conduction electron effective mass~\cite{Maisuradze2009}. 

PrPt$_4$Ge$_{12}$ has been shown to be an unconventional superconductor with two Fermi surfaces having one nodal and another nodeless gap \cite{Singh2016}. In particular, the substitution of Pr by Ce leads to the suppression of $T_c$ with the nodal gap being gradually suppressed \cite{Singh2016}. On the other hand, EuPt$_4$Ge$_{12}$ orders antiferromagnetically \cite{Ren2008,Midya2016} with a fairly large magnetic moment corresponding to a total angular momentum ($J=S=7/2$) and a Neel temperature  $T_N=1.78$ K \cite{Nicklas2011}.  For the purposes of the present work, it is important to mention that the heat capacity ($C$) measurements of Eu- or Gd-containing samples have revealed a low temperature $T$ upturn in $C/T$ as a function of temperature, which has been attributed to a Schottky anomaly resulting from the splitting of the ground state octet of Eu/Gd due to the internal molecular and external applied fields \cite{Naugle2006,Nigel2003,Kohler2006}. In contrast, the upturn in $C/T$ vs. $T$ data in samples containing Pr have been attributed to the crystalline electric field (CEF) splitting of the ground state of Pr \cite{Frederick2005, Takeda2000}.
   
As already mentioned above, the end compounds of the series Pr$_{1-x}$Eu$_x$Pt$_4$Ge$_{12}$ are superconducting ($x=0$) and antiferromagnetic \linebreak ($x=1$ respectively \citep{Nicklas2011,Singh2016}. This fact alone naturally sets the stage for investigating the interplay between  superconductivity and magnetism in this series of compounds \cite{Ijeon2017}. In this paper, we report the results and analysis of the low-temperature specific heat  measurements on Pr$_{1-x}$Eu$_x$Pt$_4$Ge$_{12}$ alloys subject to an external magnetic field. Our systematic analysis reveals that the Schottky anomaly present at low temperatures in the heat capacity is a result of the  energy-level splitting of the 
ground state octet of Eu$^{2+}$ due to an internal magnetic field. We show that this internal magnetic field is produced by the net magnetic moment due to the short-range antiferromagnetic correlations between Eu magnetic moments that coexist with superconductivity. Finally, the suppression of the superconducting critical temperature with Eu concentration can be understood using the standard tools developed for disordered multiband superconductors.
 
\section{Experimental Details}
Polycrystalline  samples  of  Pr$_{1-x}$Eu$_x$Pt$_4$Ge$_{12}$ were synthesized by arc-melting and annealing according to the procedure described in detail in Ref. \cite{Ijeon2017}. The crystal structure was determined by x-ray powder diffraction using a Bruker D8 Discover x-ray diffractometer with Cu-K$_{\alpha}$ radiation  \cite{Huang2014}. The x-ray diffraction patterns revealed that the polycrystalline samples used in this study consisted of single phase.
In order to improve the contact between the sample and the specific heat platform, the two surfaces of each sample were polished with sand paper. 
We performed a series of specific heat measurements on these polycrystalline samples of Pr$_{1-x}$Eu$_x$Pt$_4$Ge$_{12}$ ($x =$ 0, 0.05, 0.10, 0.15, 0.20,  0.30,  0.38,  0.50, 0.70, 0.80, 0.90 and 1) in zero magnetic field  and in magnetic fields $H$ up to 14 T over the temperature $T$ range $0.50$ K $\leq T \leq 15$ K. The specific  heat  $C$ measurements  were  performed  via a standard thermal relaxation technique using the He-3 option of a Quantum Design Physical Property Measurement System (PPMS).
\section{Results and Discussion}
In the following, we present our  heat capacity data for samples with various concentrations of europium ions. 
\begin{figure}
\centering
\includegraphics[width=1.0\linewidth]{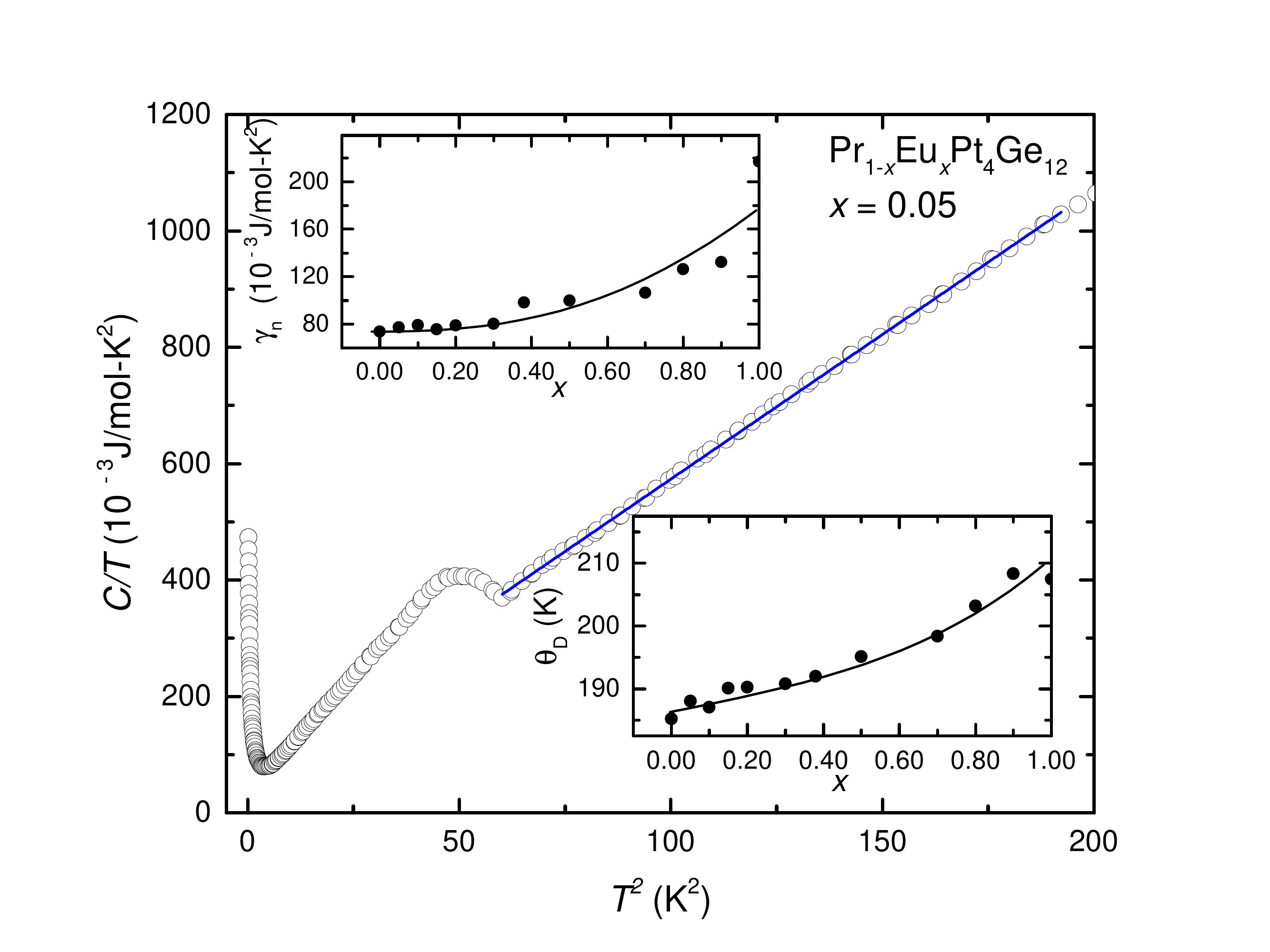}
\caption{(Color online) Specific heat $C$ divided by temperature $T$ vs. $T^2$ for the $x=0.05$ sample of Pr$_{1-x}$Eu$_x$Pt$_4$Ge$_{12}$. The solid blue line is a straight line fit of the data using the Debye model. Top left inset: Sommerfeld coefficient $\gamma$ plotted as a function of Eu concentration $x$. The solid line is a guide to the eye.  Bottom right inset: Debye temperature $\theta_D$ obtained from linear fits of the normal-state data of the main panel  plotted as a function of  $x$. The solid line is a guide to the eye.}
\label{Fig1}
\end{figure}
In Fig. ~\ref{Fig1}, we show  $C/T$ vs. $T^2$ data for the $x=0.05$ sample of Pr$_{1-x}$Eu$_x$Pt$_4$Ge$_{12}$.  In the absence of any magnetic contribution, the measured specific heat in the normal state is the sum of electronic 
$C_{\rm e}\equiv \gamma_n T$ ($\gamma_n$ is the normal-state Sommerfeld coefficient) and phonon $C_{\textrm{ph}} = \beta T^3$ contributions; hence, we fitted the measured specific heat in the normal state ($T_c < T \leq $ 15 K) for different Eu concentrations with $C(T)$ = $\gamma_n T$+$\beta T^3$. The result of such a fit for the  $x = 0.05$ sample is shown in main panel of  Fig.~\ref{Fig1} and gives ${\gamma}_n$ = 76.39$\pm$1.15mJ/mol$\cdot$K$^2$ and $\beta$ = 4.97$\pm$ 0.01mJ/mol$\cdot$K$^4$. 

The upper left inset in Fig.~\ref{Fig1} displays the dependence of the Sommerfeld coefficient $\gamma_n$ on $x$, while the bottom right inset shows the Debye temperature ($\theta_D$) as a function of x over the entire Eu concentration  range ($0 \leq x \leq 1$). We note that we obtained the Debye temperature from ${\Theta_D} = \left({12\pi^4N_Ak_B}/{5\beta}\right)^{1/3}$,
where $\beta$ is found by fitting the data, as discussed above.
Notice that both $\gamma_n$ and $\theta_D$  increase slowly with increasing Eu concentration. The $\gamma_n$ values increase from 74 mJ/mol-$K^2$ at $x=0$ to $\sim$224 mJ/mol-$K^2$ at $x=1$.

\begin{figure}
\centering
\includegraphics[width=1\linewidth]{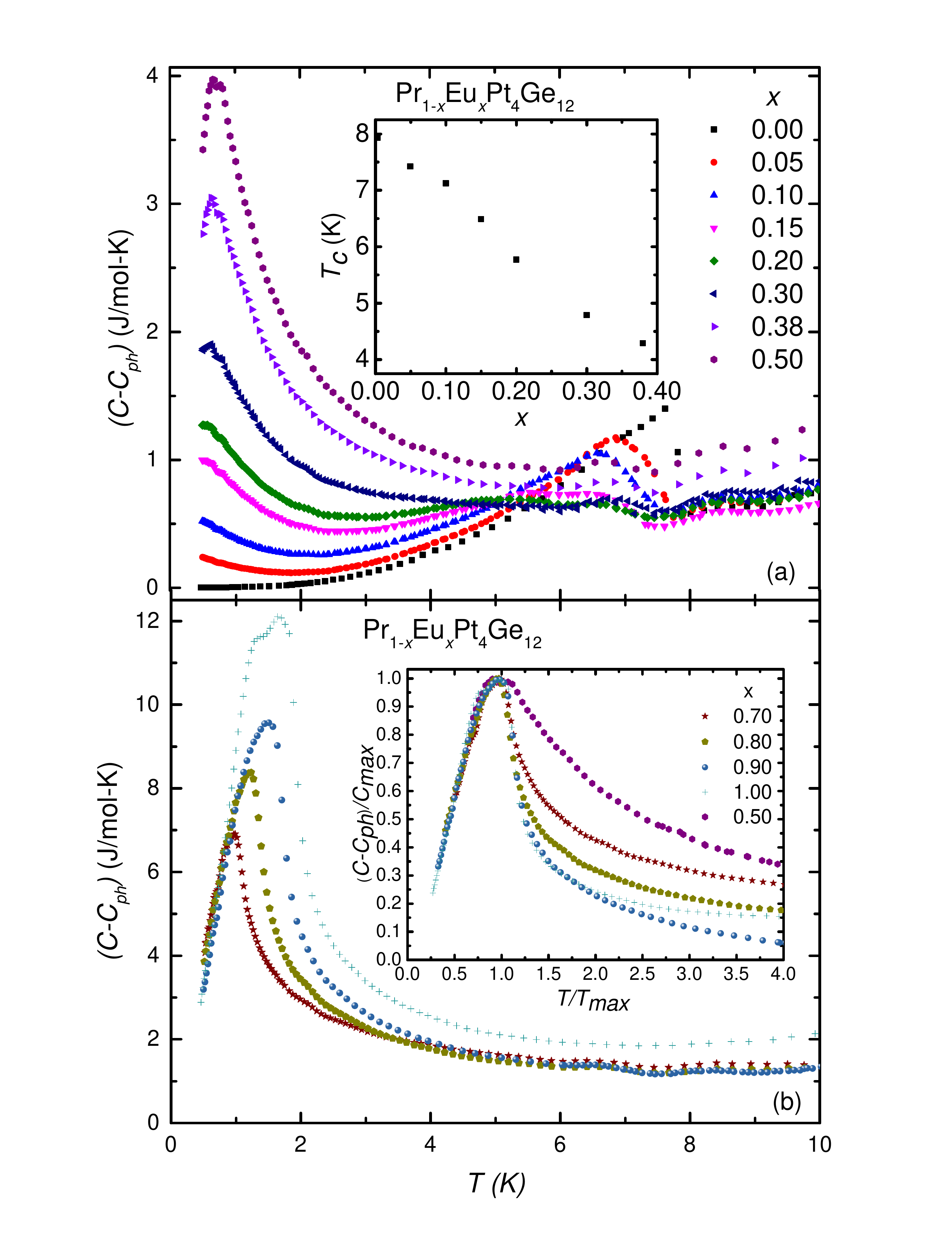}
\caption{ (Color online)  Specific heat $C-C_{ph}$ vs. temperature $T$ data for the samples  that have superconducting transitions (top) and  antiferromagnetic transitions (bottom). Top inset: Superconducting transition temperature $T_c$ as a function of Eu concentration $x$. Bottom inset: Data of the main panel normalized by $C_{max}$ and $T_{max}$, the value of the heat capacity $C-C_{ph}$ and temperature $T$ at the maximum, respectively.}
\label{Fig2}
\end{figure}

\begin{figure}
\centering
\includegraphics[width=\linewidth]{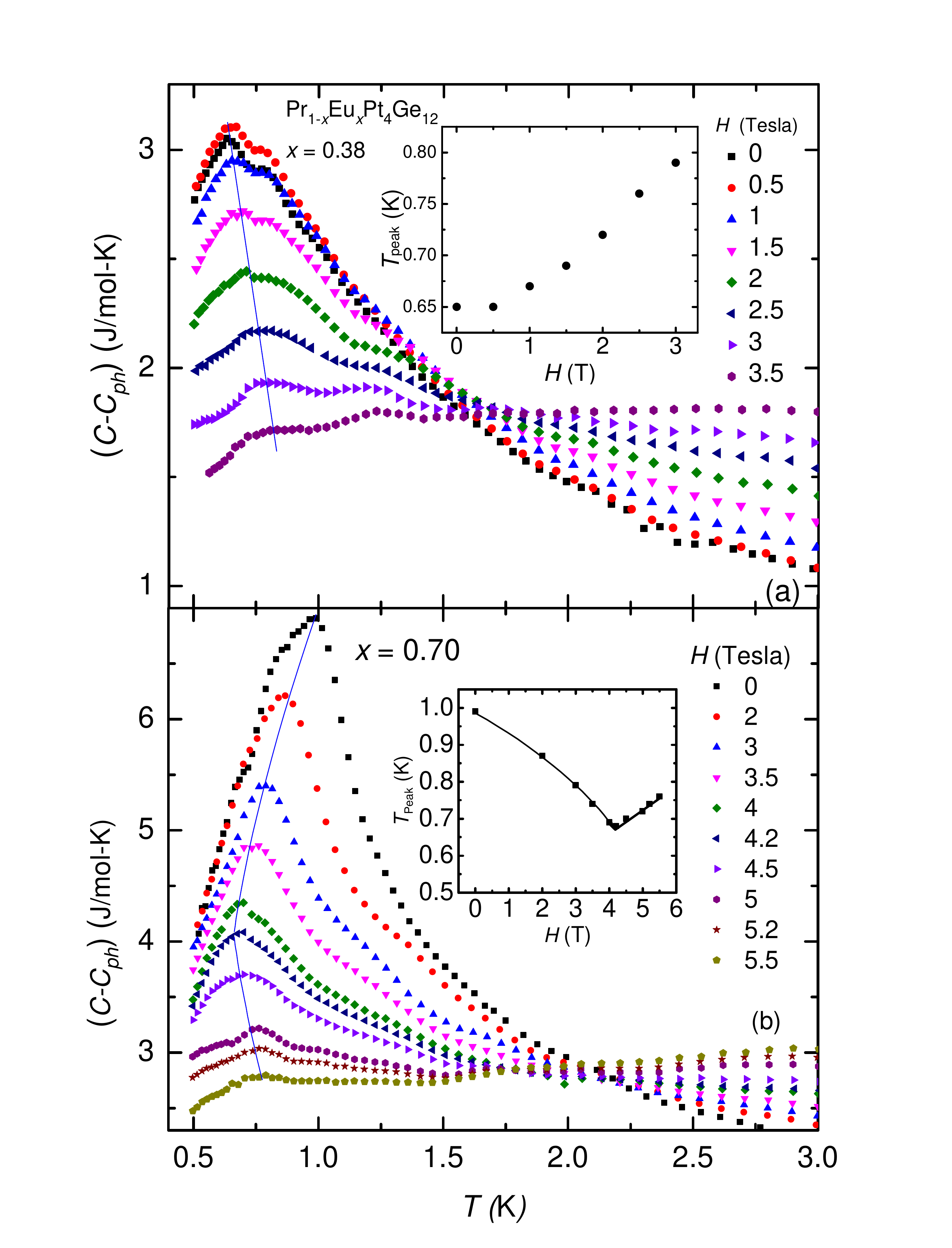}
\caption{(Color online) Heat capacity $C-C_{ph}$ vs. temperature $T$ data for Pr$_{1-x}$Eu$_x$Pt$_4$Ge$_{12}$ with (a) $x=0.38$ and (b) $x=0.7$, measured in various applied magnetic fields. The solid lines through the peaks of the HC curves are guides   to the eye.Insets: Temperature $T_{peak}$, corresponding to the maximum in heat capacity data of the main panel, plotted as a function of magnetic field $H$.}
\label{Fig3}
\end{figure}

We present in Fig.~\ref{Fig2}(a) specific heat $C-C_{ph}$ data as a function of $T$ obtained after subtracting the phonon contribution (obtained from the above fit) from the measured specific heat
of the samples that display superconducting transition in heat capacity and resistivity measurements.  A superconducting transition is clearly observed for Pr$_{1-x}$Eu$_x$Pt$_4$Ge$_{12}$ samples with $x \leq 0.3$ in heat capacity measurements, whereas this phase transition is observed in the resistivity measurements for the alloys with $x\leq0.5$ \cite{Ijeon2017}. 
Furthermore, the inset to Fig.~\ref{Fig2}(a) displays the suppression of the superconducting transition temperature $T_c$ as a function of $x$.  Samples with $x >$ 0.5 do not display a superconducting transition for temperatures down to \linebreak $T=0.5$ K.

It has been shown that EuPt$_4$Ge$_{12}$ displays an antiferromagnetic (AFM) transition at  $T_N=1.77$ K \cite{Nicklas2011,Grytsiv2008}, manifested by a peak in $C(T)$ at this $T=T_N$. Therefore, we conclude that the peaks in $C(T)$ of the samples with $x>0.5$ [see Fig.~\ref{Fig2}(b)] must account for the second-order phase transition into an AFM state.  The  inset to this figure shows that, indeed, the $C(T)$ data normalized to the corresponding values of the peak  scale for the $x>0.5$ samples and do not scale for the $x=0.5$ sample. This reveals that  the peak in $C(T)$ for $x>0.5$ has the same origin, namely, AFM order.
   
We now turn our attention to the low-temperature region of the $C-C_{ph}$ vs. $T$  data shown in Fig.~\ref{Fig2}(a). All these Eu substituted samples have systematic  upturns in the low-temperature region. The samples with $x>0.15$ show a maximum in the electronic specific heat that shifts to higher temperature and increases in magnitude with increasing Eu  content. For the case of the samples with $x\leq0.15$, the electronic specific heat curves show an upturn, without reaching a maximum in the measured temperature range down to 0.5 K.
We attribute this anomaly  to a Schottky-type anomaly that arises from the energy-level splitting of the ground state octet of Eu$^{2+}$. We note that, generally, one expects the crystalline electric fields (CEF) to lift the spherical $(2J+1)$-fold degeneracy of the ground state of rare-earth ions due to the lower symmetry of the crystalline environment by coupling of the electric fields to the orbital degrees of freedom. The resulting multiplet structure depends on the strength of the crystalline electric field and the symmetry of the local  rare-earth environment.  The ground state of Eu$^{2+}$ is 4$f^7$ and, hence, the Hund's rule ground state is $^8S_{7/2}$ with  $L=0$, so that the crystalline fields cannot lift the \linebreak {$(2J+1)$-fold} degeneracy. In addition, the Schottky anomaly peaks in $C(T)$ cannot reflect the CEF splitting of the ground state of Pr$^{3+}$ since the energy required to lift the degeneracy of the ground-state is higher for Pr$^{3+}$  \cite{Frederick2005}, hence the Schottky peaks appear at higher $T$ ($\sim 10$ K) than we presently observed ($T < 1$ K). Therefore, the Schottky anomaly must be due to \emph{an internal magnetic field}. Interestingly, a similar effect has  been observed in RuSr$_{2}$(Gd$_{1.5}$Ce$_{0.5}$)Cu$_2$O$_{10-\delta}$, YbPd$_2$Sn \cite{Aoki2000} and Yb$_{0.24}$Sn$_{0.76}$Ru \cite{Thompson2011}. For example, in RuSr$_{2}$(Gd$_{1.5}$Ce$_{0.5}$)Cu$_2$O$_{10-\delta}$ alloys this effect has been attributed to the lifting of the degeneracy of the ground state $^8S_{7/2}$ of Gd$^{3+}$ by internal and external magnetic fields \cite{Naugle2006}: Gd and Eu have the same ground state with $L=0$, so the splitting of the degenerate Eu ground state in Pr$_{1-x}$Eu$_x$Pt$_4$Ge$_{12}$ may also be a result of an internal magnetic field that  coexists with superconductivity. As we will show below, the origin of this internal magnetic field is very likely due to short-range antiferromagnetic correlations between the Eu atoms.

In order to confirm that this is indeed the case, we measured the heat capacity of all the samples studied in various applied magnetic fields. The results for $C-C_{ph}$ vs $T$ for the $x=0.38$ sample are shown in Fig.~\ref{Fig3}(a) and for the $x=0.70$ sample in Fig.~\ref{Fig3}(b). Our results can be summarized as follows: with increasing magnetic field,  the peak in the specific heat of the samples  with $0.2\leq x\leq0.50$ shifts to higher temperatures  [Fig.~\ref{Fig3}(a)], while the peak of the higher Eu doped samples ($x>0.5$) first shifts to lower temperatures  and then to higher temperatures with further increasing $H$ [Fig.~\ref{Fig3}(b) and its inset]. The initial shift of the peak in $C(T)$ to lower temperatures with increasing magnetic field in these higher Eu concentration samples is a result of the suppression of the AFM transition by applied field. Once the AFM transition is suppressed, the peak in the specific heat shifts to the higher temperatures with further increase of $H$, as observed in the samples with $x \leq 0.50$. We note that the upturn in specific heat  in the samples with $0.05\leq x\leq 0.15$ starts to show a clear peak with increasing $H$, confirming that the upturn observed in these lower Eu concentration samples is also a result of the splitting of the degenerate ground state of Eu. Therefore, these results show that, indeed, the Schottky anomaly peaks revealed in $C(T)$ at low temperatures are a result of the splitting of the degenerate ground state of Eu by internal and/or external magnetic fields. 

\begin{figure}
\centering
\includegraphics[width=1\linewidth]{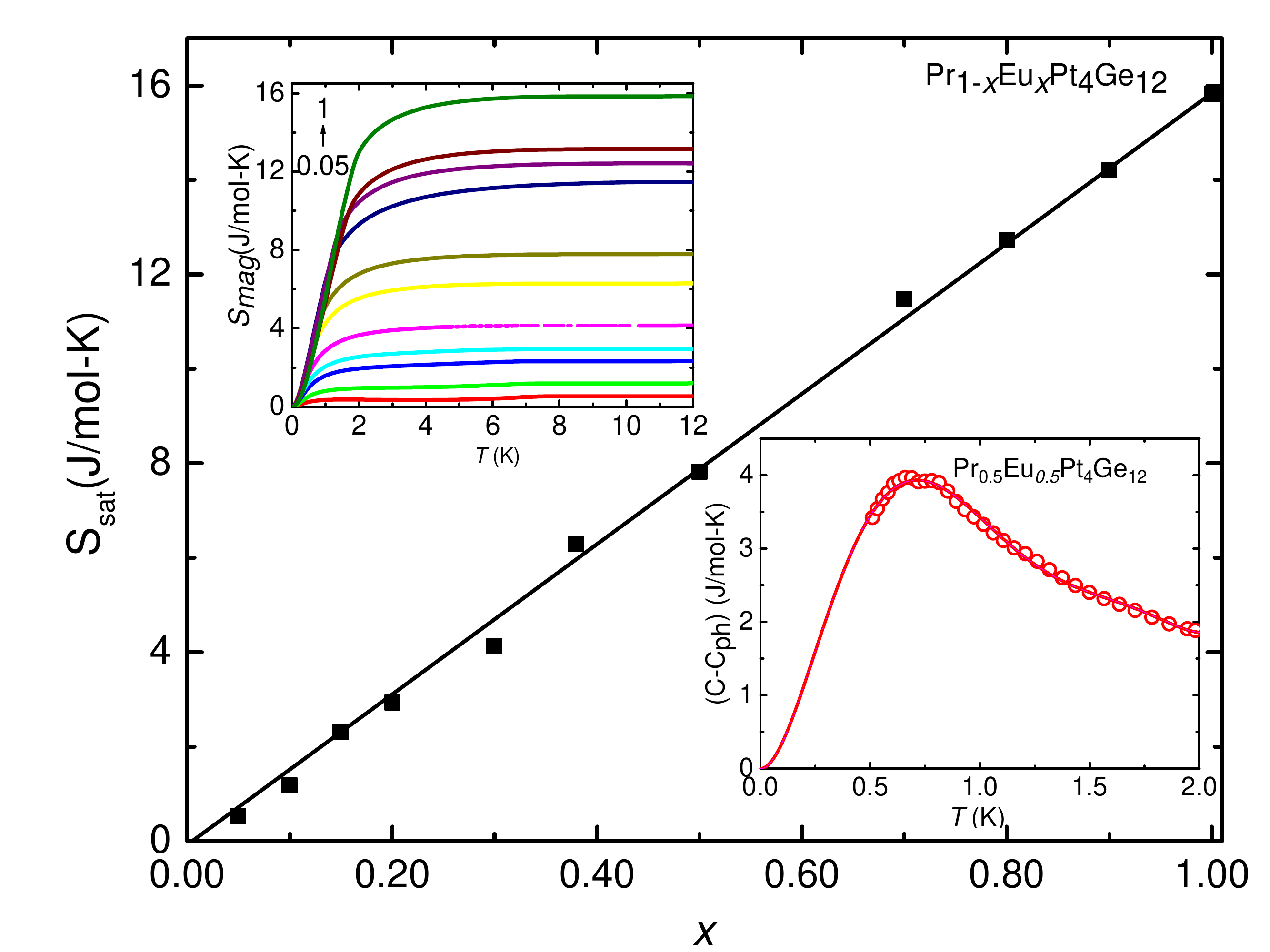}
\caption{(Color online)  Saturation entropy $S_{\textrm{sat}}$ plotted as  function of Eu concentration $x$ over the whole measured  range($0\leq x \leq1$). The straight line is a fit of $S_{\textrm{sat}}$ vs $x$ with $S = a + bx$ where $a = (-0.29\pm 0.14) $J/mol-K and   $b = (16.26\pm2.42)$ J/mol-K. Top inset:  Magnetic entropy $S_{\textrm{mag}}$ vs. temperature $T$. Bottom inset: Extrapolation of the $C-C_{ph}$ vs $T$ data  to 0 K using a polynomial fit for the $x=0.50$ sample.}
\label{Fig4}
\end{figure}

Further evidence that the internal magnetic field responsible for the observed Schottky anomaly is produced by antiferromagnetically-correlated Eu ions can be found by calculating the magnetic entropy $S_{\textrm{mag}}$. To extract the entropy from our low-$T$ heat capacity measurements, we first estimated the $C-C_{ph}$ vs. $T$  curve down to 0 K  by fitting the low temperature data and extrapolating this fit  to 0 K (see bottom inset of Fig. \ref{Fig4}). 
We then calculated the magnetic entropy 
$S_{\textrm{mag}}= \int_0^{T}C_{\textrm{mag}}(T)dT/T$,
where $C_{\textrm{mag}}\equiv C-C_{ph}-\gamma_n T$. We plot the results we obtained for the magnetic entropy at different temperatures and for different Eu-substituted samples in the inset of Fig.~\ref{Fig4}. Note that the magnetic entropy first increases with increasing temperature and then it saturates (see inset to Fig. 4). The saturation value of the entropy corresponds to the magnetic entropy when all eight levels are occupied.
  
The main panel of Fig.~\ref{Fig4} is a plot of the saturation value $S_{\textrm{sat}}$ of the magnetic entropy at $T=10$ K vs. $x$. The fact that the internal magnetic field is produced by antiferromagnetically correlated Eu ions is also supported by the fact that  the entropy increases linearly with increasing Eu concentration
. The straight line is a  linear fit of the $S_{sat}$ vs. $x$ data, which gives a slope  of 16.26$\pm$2.42 J/mol-K and an intercept of $-0.29\pm$0.14 J/mol-K.  

Furthermore, the single-ion entropy associated with a state of angular momentum $J$ is \linebreak $S=xR\ln(2J+1)$, where $x$ is Eu concentration in Pr$_{1-x}$Eu$_x$Pt$_4$Ge$_{12}$.  Hence, the corresponding entropy due to the ground-state splitting of the $^8S_{7/2}$ of Eu is $S\approx17.28 x$ J/mol-K. Note that this value is within 6$\%$ of the value for the magnetic entropy (16.26$\pm$2.42 J/mol-K for $x=1$) obtained from the least-square-linear fit of Fig.~\ref{Fig4}. The excellent agreement between the calculated and expected entropies shows that all eight energy levels of Eu are occupied and that the errors in determining the phonon and electronic contributions to  the specific heat, as well as the errors involved in the extrapolation of these quantities to 0 K are very small. Therefore, all of these results further confirm our initial assumption that short-range AFM correlations between Eu ions are mainly responsible for the Schottky anomaly revealed by the low-$T$ specific heat measurements. 
  
Thus, we have established that the low temperature upturn in $C-C_{ph}$ of Pr$_{1-x}$Eu$_x$Pt$_4$Ge$_{12}$ as $T$ decreases is caused by the splitting of the degenerate  $^8S_{7/2}$ ground state of Eu into eight equally-spaced levels.
Recall that the Schottky heat capacity anomaly for a multilevel system with the degeneracy fully lifted by a magnetic field is given by \cite{Gopal1966}:
\begin{equation}\label{Sch}
\begin{split}
C_{\textrm{Sch}}&=r(x)\frac{R}{T^2}\left[f_2(T)/f_0(T)-f_1^2(T)/f_0^2(T)\right], \\
f_m(T)&=\sum\limits_{j=0}^7\Delta_i^m\exp\left(-\Delta_j/ k_BT\right), 
\end{split}
\end{equation}
where $\Delta_j=j\cdot\Delta$ is the energy  gap between the lowest energy level ($j=0$) and the $j^{th}$ energy level, $R=8.31$ J/mol-K is the universal gas constant, $\Delta= g\mu_BH$, $g=2$ ($L=0$), $\mu_B$ is the Bohr magnetron and $H$ is the magnetic field.  We have also included the parameter $r(x)$ which is proportional to the concentration of Eu ions and will be used as one of the fitting parameters.  Furthermore, for the analysis of the heat capacity data it is important to keep in mind that EuPt$_4$Ge$_{12}$ orders anti-ferromagnetically, so
that in the alloys Pr$_{1-x}$Eu$_x$Pt$_4$Ge$_{12}$ one may expect short-range antiferromagnetic correlations between the Eu ions. Therefore, we hypothesize that short-ranges antiferromagnetic correlations between Eu ions are present even in samples with small $x$. One consequence of the short-range antiferromagnetic correlations is that the nearest neighbor Eu ions
will provide a net magnetic moment which should lift the $(2J+1)$-degeneracy of the Eu ion. The level splitting on each Eu ion will of course be different due to the randomness associated with the alloying itself. Therefore, for our analysis of the heat capacity data, we need to use the expression (\ref{Sch}) averaged over the distribution of the level splittings $\Delta$. To perform the averaging, we assume that the level splitting energy $\Delta$ is evenly distributed within a certain interval of values $\Delta\in[\Delta_{\textrm{min}},\Delta_{\textrm{max}}]$ around the mean value $\overline{\Delta}=(\Delta_{\textrm{min}}+\Delta_{\textrm{max}})/2$ \cite{Aoki2000}.

\begin{figure}
\centering
\includegraphics[width=1\linewidth]{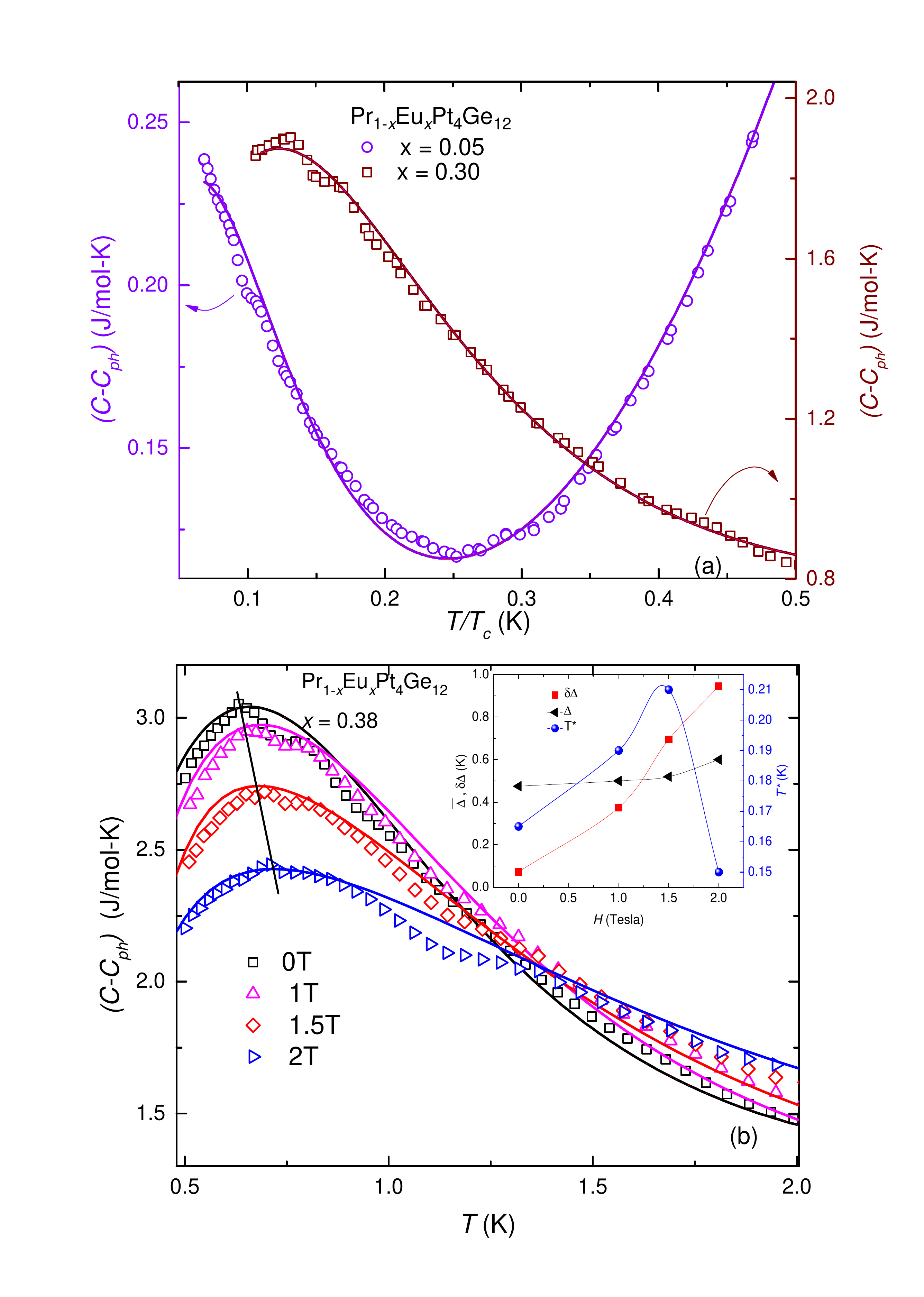}
\caption{(Color online) (a) Fit of the heat capacity $C-C_{ph}$ vs $T/T_c$ with the Schottky and superconducting contributions for $x=0.05$ (left axis) and $x=0.30$ (right axis) samples. See text for details.  (b) Fit of the field dependence of the heat capacity with a distribution of internal field as described in the text. Inset to (b): Fitting parameters as a function of external magnetic field.
\label{Fig5} }
\end{figure}

\begin{figure}
\centering
\includegraphics[width=\linewidth]{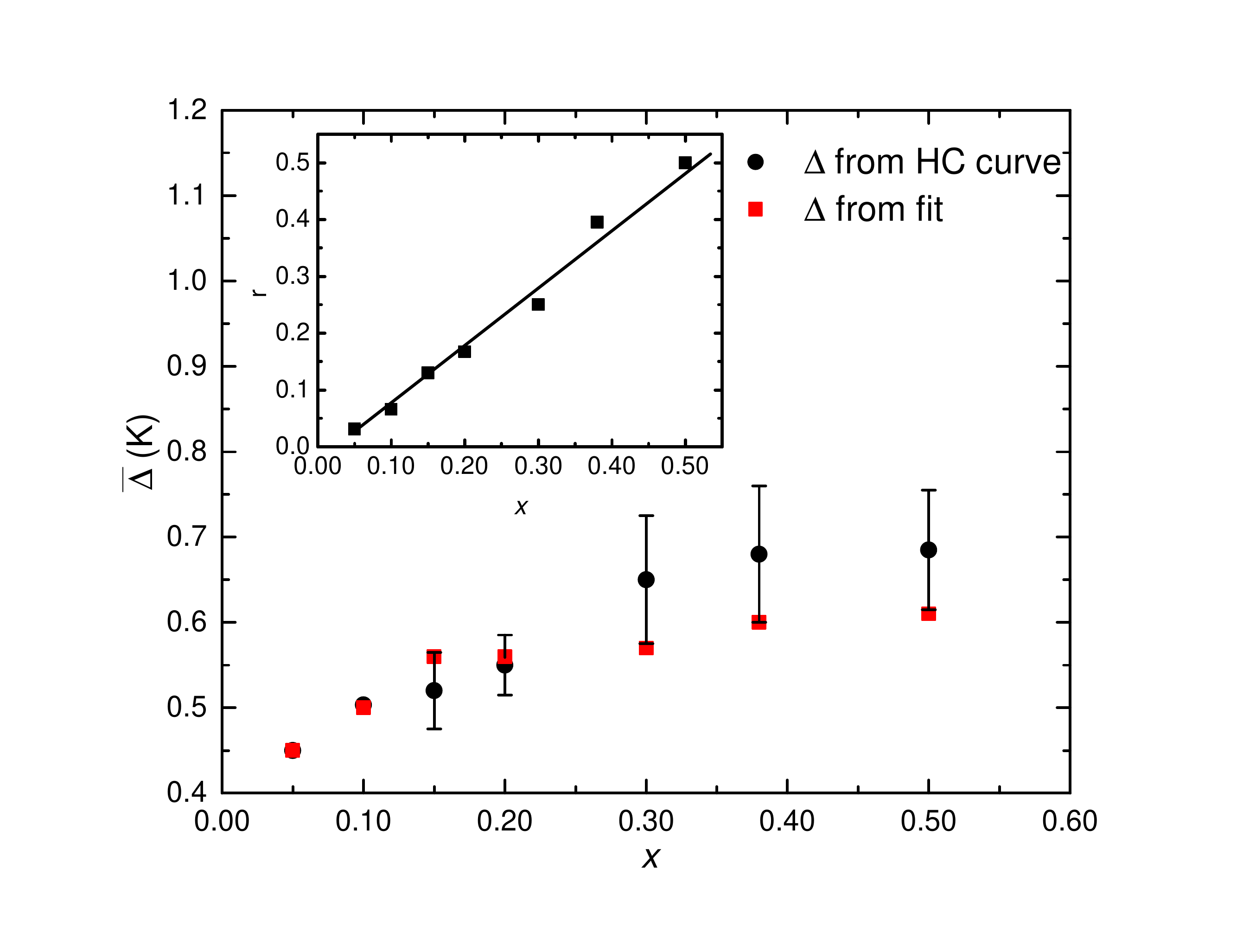}
\caption{(Color online) Energy gap $\Delta$ vs Eu concentration $x$. The black solid circles are $\Delta$ extracted from the temperature at which the heat capacity peaks occur in the plot of $C-C_{ph}$ vs $T$ curves; the red squares are the energy gaps obtained by fitting the $C-C_{ph}$ vs $T$ data . Inset:  Coefficient $r$ obtained from the fit plotted as a function of $x$. The line is a straight line fit with $x = r$.}
\label{Fig6}
\end{figure}

Our results show that for the case of zero magnetic field we can use expression (\ref{Sch}) directly with $\Delta=\overline{\Delta}$. For  $x \leq 0.10$, we fitted the specific heat data below the temperature where the upturn starts with a the sum of  the Schottky contribution given by (\ref{Sch}) and a superconducting nodal contribution of the form $C_{\textrm{SC}}=AT^n$, where $A$ and $n$ are fitting parameters, since a better fit of the HC data is obtained with a nodal gap. For the doping range $x>0.10$, a better fit of  the HC data is obtained with the sum of Eq. (\ref{Sch}) and the superconducting contribution  with an isotropic gap $C_{\textrm{SC}}=B\exp^{-\delta/T}$ with $B$ and $\delta$ as fitting parameters; this is perfectly expected since the nodal gap is quickly suppressed by scattering on lattice imperfections. 
In Fig. ~\ref{Fig5}(a), we show the results of the fit of the heat capacity data for the $x=0.05$ and $x=0.30$ samples.

When the external magnetic field is applied, the peak in the heat capacity shifts to higher temperatures, Fig. \ref{Fig3}. This feature can 
be explained by the fact that the application of the field leads to broadening in the distribution of the energy splittings. As a result the averaging of the heat capacity over the probability distribution discussed above leads to an overall decrease in the amplitude
of the peak together with its broadening \cite{Aoki2000}. In Fig. ~\ref{Fig5}(b), we show the fits to the heat capacity data for a sample with $x=0.38$ which again include the Schottky contribution averaged over $\Delta$ and the superconducting contribution with an isotropic gap, as discussed above. For the fits of the Schottky peak we actually took into account that the internal magnetic field, $\overline{\Delta}$ is weakly temperature dependent. Specifically, we found that the best fits are obtained by replacing $\overline{\Delta}\to \left(\frac{T}{T-T^*}\right)\overline{\Delta}$
and using $T^*$ as a fitting parameter. The field dependences of $T^*$, $\overline{\Delta}$, and $\delta\Delta \equiv\Delta_{\textrm{max}}-\Delta_{\textrm{min}}$ for the results presented in Fig. ~\ref{Fig5}(b) are shown in its inset. 

The results for the average energy splitting $\overline{\Delta}$ and coefficient $r(x)$ obtained from the fits of the low-temperature HC data measured in zero external magnetic field are presented in Fig.~\ref{Fig6}. The temperature corresponding to the Schottky peak for a system with eight energy levels and the degeneracy fully lifted yield $\overline{\Delta}$, i.e., $T_{\textrm{peak}}$ = $\overline{\Delta}$. The black circles in the main panel give the values of $\overline{\Delta}$ obtained directly from $T_{\textrm{peak}}$ on the HC curves, while the red squares give the values of $\overline{\Delta}$ obtained from the fits of the specific heat curves as discussed above. It is noteworthy that the values of $\overline{\Delta}$ obtained through these two procedures are in excellent  agreement. This represents a further confirmation that the eight energy levels of Eu are mainly responsible for the observed  HC anomaly. The inset to Fig. ~\ref{Fig6} shows the coefficient $r$ in Eq. (\ref{Sch}) which accounts for the percentage of Eu in Pr$_{1-x}$Eu$_x$Pt$_4$Ge$_{12}$. The straight line in the inset is a fit of the $C$ vs. $x$ data with slope and intercept close to 1 and 0, respectively.

\begin{figure}
\centering
\includegraphics[width=1\linewidth]{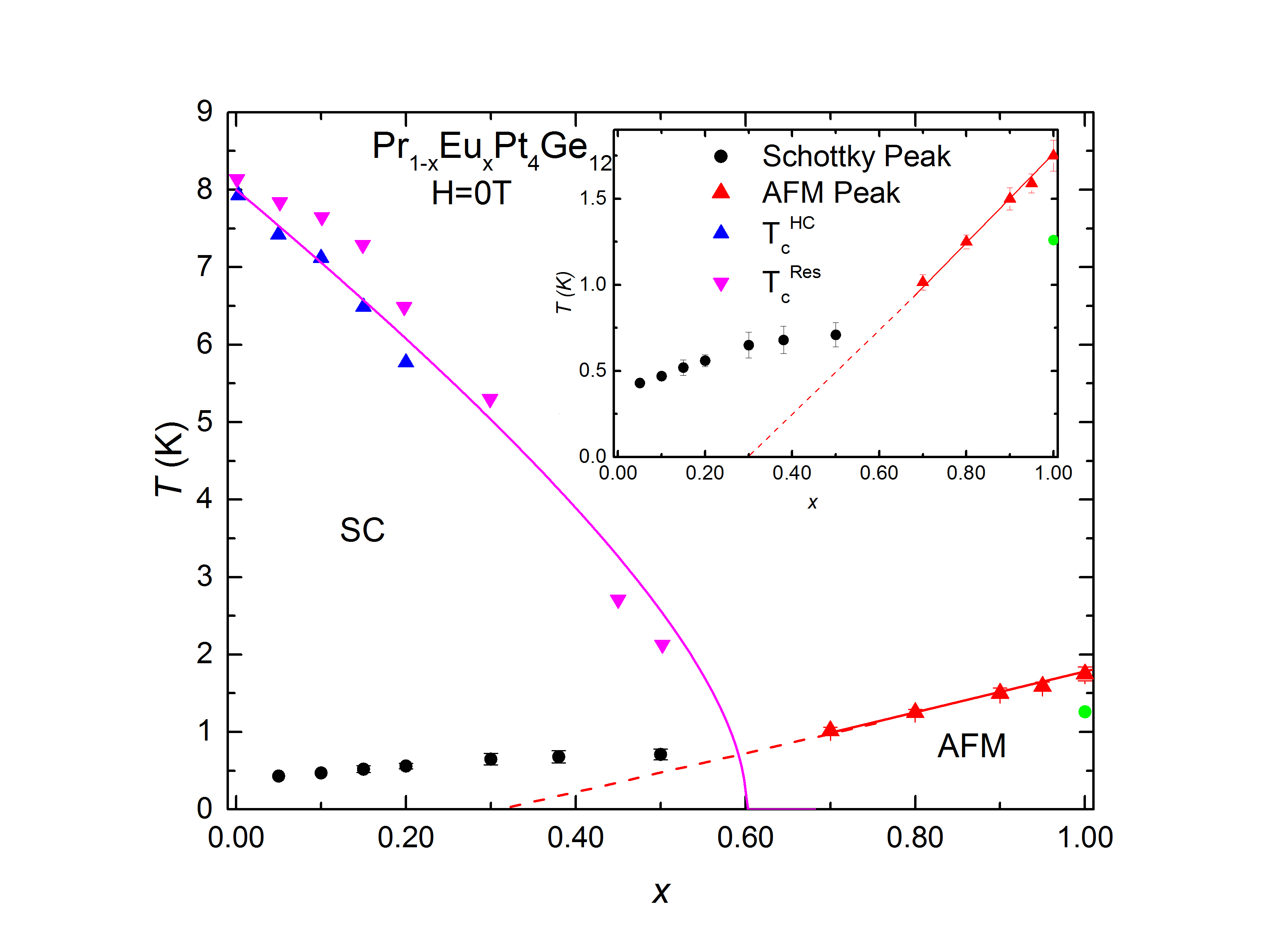}
\caption{(Color online) Plot of temperature $T$ vs. Eu concentration $x$ phase diagram. Pink triangles are the superconducting transition temperature $T_c$ obtained from HC and resistivity measurements. The pink solid line is a theoretical fit to the data using expression (\ref{Tcx}). The black solid circles give the temperature of the Schottky peak, while the red triangles are the AFM transition temperature extracted from the HC curves. The solid red line is a guide to the eye and the dashed red line is the extrapolation of the AFM phase boundary to lower temperatures. The solid green point is the temperature where the second peak appears in the $x=1$ sample. Inset: Zoomed phase diagram of the data in the main panel to better reveal the values of the  Schottky and AFM peaks.  
\label{Fig7} }
\end{figure}

Finally, in Figure~\ref{Fig7} we show the temperature $T$ vs. Eu concentration $x$ phase diagram of Pr$_{1-x}$Eu$_x$Pt$_4$Ge$_{12}$. The $T_c(x)$ data were obtained from heat capacity (blue triangles) and resistivity (pink triangles) measurements. Note that $T_c(x)$ gradually decreases, with a negative curvature, with increasing $x$ up to $x=0.5$. The solid pink curve is a theoretical fit to the data computed using the following expression:
\begin{equation}\label{Tcx}
\ln\left(\frac{T_{c0}}{T_c}\right)=\psi\left(\frac{1}{2}+\frac{\Gamma}{\pi T_c}\right)-\psi\left(\frac{1}{2}\right).
\end{equation} 
Here $\psi(z)$ is the digamma function, $T_{c0}=T_c(x=0)$ and $\Gamma\propto x$ is the disorder induced single-particle scattering rate. We found that expression (\ref{Tcx}) describes the experimental data best when the critical value of the scattering rate for which $T_c\to 0$ is $\Gamma_c\approx 0.73T_{c0}$. Equation (\ref{Tcx}) is similar to the one describing the suppression of the critical temperature in conventional $s$-wave superconductors with magnetic impurities \cite{AGD}, multiband $s^{\pm}$ superconductors with potential impurities with $\Gamma$ being the interband scattering rate \cite{VC_Disorder2011,London2015}, and last, but not least, $d$-wave superconductors with nonmagnetic impurities \cite{Vekhter2006,Vitya2008,Chubukov2010}. The expression (\ref{Tcx}) describes the data quite well which is not surprising in light of the facts that the superconducting order parameter on one of the bands is nodal, while the pairing amplitudes on the other bands are nodeless \cite{Singh2016}. 

In Fig. \ref{Fig7} and its inset, the solid black circles represent the temperatures where the Schottky peaks occur. The red triangles are the AFM transition temperatures corresponding to different $x$ values. Clearly, the plot of the Schottky and AFM peak temperatures vs. $x$ have two slopes (see inset  to Fig.~\ref{Fig7}), indicating the different origin of the two peaks present in $C_e(T)$.  The present study shows that there is a coexistence of superconductivity and antiferromagnetically correlated Eu ions for $x \leq 0.5$ and that for $0.3\leq x \leq 0.6$ there may  be a coexistence of SC and long-range AFM. Further lower temperatures studies are required to address this second region of the phase diagram.

The compound EuPt$_4$Ge$_{12}$ has been reported to show at least three peaks in $C_e(T)$ close to the AFM transition \cite{Nicklas2011}. In the present study we observed two peaks in $C_e(T)$: one at the AFM transition $T_N=1.78$ K and another one at $T=1.2$ K $< T_N$. We measured the heat capacity of this compound in magnetic field in order to determine the response of the two peaks to magnetic field. We observed that the AFM peak shifts to lower temperatures and its amplitude decreases with increasing $H$, whereas the second peak is suppressed with increasing magnetic field and no longer appears in higher field. We were unable to resolve such two peak feature for other lower doped anti-ferromagnetic samples through HC measurements.

\section{Conclusions}
We analyzed the low-temperature specific heat data in order to investigate the effect of Eu substitution on the nature of the superconducting and antiferromagnetic orders in the Pr$_{1-x}$Eu$_x$Pt$_4$Ge$_{12}$ filled skutterudite system. The superconducting transition temperature is monothonically suppressed with increasing Eu concentration. Our data reveal the presence of short AFM correlations between Eu ions under the superconducting dome for $ x \leq 0.50$. These AFM correlations produce a local internal magnetic field, which lifts the eight-fold degeneracy of the Eu ground state and gives rise to a Schottky peak in heat capacity. The superconducting gap of Pr$_{1-x}$Eu$_x$Pt$_4$Ge$_{12}$ has line nodes, i.e., $C_{SC}\propto T^ 2$ for the doping range $0\leq x\leq 0.10$ and it is isotropic, i.e., $C_{SC} \propto e^{-\delta/T}$ for $ 0.15\leq x \leq 0.50$.  This system displays long-rage AFM order for $x\geq0.70$, with the AFM transition temperature decreasing with increasing Pr concentration. Antiferromagnetism and superconductivity most likely coexist for $0.30\leq x \leq0.60$. 

\section{Acknowledgments} 
This work was supported by the National Science Foundation grants DMR-1505826 and DMR-1506547 at KSU and by the
US Department of Energy, Office of Basic Energy Sciences, Division of Materials Sciences and Engineering, under Grant
No. DE-FG02-04ER46105 at UCSD. The work of M.D. was financially supported in part by the U.S. Department of Energy, Office of Basic Energy 
Sciences under Award No. DE-SC0016481. 
\bibliography{Ref_skutt}

\end{document}